\documentclass[conference]{IEEEtran}
\IEEEoverridecommandlockouts
\usepackage{cite}
\usepackage{amsmath,amssymb,amsfonts}
\usepackage{amsthm}
\usepackage{graphicx}
\usepackage{textcomp}
\usepackage{xcolor}
\usepackage{mathtools}
\usepackage{subcaption}
\usepackage{mwe}
\usepackage[ruled,vlined]{algorithm2e}
\usepackage{footnote}
\usepackage[skip=8pt, font=small]{caption}
\DeclareMathOperator*{\argmin}{arg\,min}

\DeclareRobustCommand{\mcal}[1]{%
  \ifcat\noexpand#1\relax\mathnormal{#1}\else\cal{#1}\fi
}
\DeclareRobustCommand{\BM}[1]{%
  \ifcat\noexpand#1\relax\bm{\boldUppercaseItalicGreek{#1}}\else\bm{#1}\fi
}
\makeatletter
\newcommand{\boldUppercaseItalicGreek}[1]{\csname var\expandafter\@gobble\string#1\endcsname}
\makeatother

\newcommand\given[1][]{\:#1\vert\:}

\renewcommand{\IEEEQED}{\IEEEQEDopen}

\usepackage{bm}
\usepackage{color, courier}
\usepackage{psfrag}
\usepackage{acronym}
\usepackage{amsmath}
\usepackage{amssymb}
\usepackage{amsfonts}
\usepackage{latexsym}
\usepackage{mathrsfs}
\usepackage{epstopdf}
\usepackage{mathtools}
\usepackage[caption=false, font=scriptsize]{subfig}
\usepackage{relsize}
\usepackage{comment}
\usepackage{authblk}

\definecolor{green}{rgb}{0, 0.5, 0}
\definecolor{pink}{rgb}{1, 0, 1}

\def\BibTeX{{\rm B\kern-.05em{\sc i\kern-.025em b}\kern-.08em
    T\kern-.1667em\lower.7ex\hbox{E}\kern-.125emX}}
\begin{document}

\title{
A Constrained RL Approach for Cost-Efficient Delivery of Latency-Sensitive Applications 
\thanks{This material is based upon work supported in part by NYU Wireless, the National Science Foundation under grant no. 21482148315, and funds from federal agency and industry partners as specified in the Resilient \& Intelligent NextG Systems (RINGS) program.}
}
\author[1]{Ozan Ayg{\"u}n}
\author[2]{Vincenzo Norman Vitale}
\author[3]{Antonia M. Tulino}
\author[4]{Hao Feng}
\author[5]{Elza Erkip}
\author[6]{Jaime Llorca}
\affil[1,3,5,6]{New York University, NY, USA\\Email: \{ozan, atulino, elza, jllorca\}@nyu.edu}
\affil[2,3]{Università degli Studi Napoli Federico II, Naples, Italy\\
Email: vincenzonorman.vitale@unina.it}
\affil[4]{Intel Labs\\ Email: haofeng.fh666@gmail.com}
\affil[6]{Università degli Studi di Trento, Trento, Italy\\ Email: jaime.llorca@unitn.it}
\affil[6]{Centre Tecnologic de Telecommunicacions de Catalunya (CTTC), Barcelona, Spain.}

\maketitle

\begin{abstract}
Next-generation networks aim to provide performance guarantees to real-time interactive services that require timely and cost-efficient packet delivery. In this context, the goal is to reliably deliver packets with strict deadlines imposed by the application while minimizing overall resource allocation cost. A large body of work has leveraged stochastic optimization techniques to design efficient dynamic routing and scheduling solutions under average delay constraints; however, these methods fall short when faced with strict per-packet delay requirements. We formulate the {\em minimum-cost delay-constrained} network control problem as a constrained Markov decision process and utilize constrained deep reinforcement learning (CDRL) techniques to effectively minimize total resource allocation cost while maintaining timely throughput above a target reliability level. Results indicate that the proposed CDRL-based solution can ensure timely packet delivery even when existing baselines fall short, and it achieves lower cost compared to other throughput-maximizing methods.

\end{abstract}

\begin{IEEEkeywords}
dynamic network control, timely throughput, cost minimization, constrained deep reinforcement learning
\end{IEEEkeywords}
\section{Introduction}

Next-generation (NextG) networks are designed to advance beyond 5G by supporting ultra-responsive and intelligent systems for real-time interactive (RTI) applications such as remote surgery, autonomous driving, and immersive virtual reality \cite{cai2022compute}. Ultra-Reliable Low-Latency Communications is essential to meet the stringent low-latency performance requirements of such RTI services, while cost-efficient resource allocation is equally critical for sustainable network operation. RTI services impose strict per-packet latency constraints to ensure timely delivery, which is vital for quality of service (QoS). Meanwhile, minimizing operational cost (e.g., power consumption) remains critical for network operators, necessitating efficient routing and scheduling strategies that carefully balance latency with resource efficiency.

The dynamic packet routing and scheduling problem with bounded average delay constraints has been widely studied, with the backpressure (BP) algorithm providing distributed, throughput-optimal policies leveraging Lyapunov drift control theory \cite{tassiulas1990stability, neely2022stochastic}. However, BP-based methods often experience high delays due to packet cycling, a challenge addressed by centralized routing and distributed scheduling algorithms like universal max-weight (UMW) and universal cloud network control (UCNC) \cite{sinha2017optimal, zhang2021optimal}. These methods reduce delays by assigning acyclic routes at packet sources, maintaining throughput optimality.

The approaches above address packet routing and scheduling with average delay constraints. However, packets in RTI services must be delivered within prescribed {\em packet lifetimes},  
or they are deemed outdated and ineffective to the application
\cite{singh2018throughput}. 
The {\em minimum-cost delay-constrained network control} (MDNC) problem was examined in \cite{cai2022ultra}, where the authors designed a heuristic solution and showcased the challenge of applying stochastic optimization based algorithms 
due to the distinct nature of lifetime-based 
queueing dynamics. Instead, this problem can be modeled as a constrained Markov decision process (MDP), allowing reinforcement learning (RL) methods to be employed for the design of efficient solutions.
RL-based techniques were proposed for packet routing and scheduling in networks modeled as MDPs \cite{boyan1993packet, mammeri2019reinforcement}. However, current RL approaches focus on queue backlog stabilization \cite{liu2022rl}, maximizing timely packet delivery \cite{hasanzadezonuzy2020reinforcement}, reducing average delivery times \cite{you2020toward}, or resource allocation \cite{liu2021constraint}. To the best of our knowledge, no existing RL methods address cost minimization while handling packet routing and scheduling with specific delivery deadlines.
Our contributions in this paper are as follows:
\begin{itemize}
    \item We show that the MDNC problem can be modeled as a constrained MDP (CMDP) where the optimal policy can be learned through a constrained deep RL (CDRL) driven dual subgradient algorithm, which we refer to as CDRL-NC,
    \item We propose a multi-agent CDRL-NC framework where a centralized routing agent and distributed scheduling agents can cooperatively learn feasible network control policies that ensure reliable timely packet delivery while minimizing the resource allocation cost,
    \item We demonstrate through dynamic event-driven network simulations that CDRL-NC achieves significantly lower resource allocation cost than state-of-the-art approaches while meeting stringent timely throughput constraints, and consistently satisfies these constraints even in scenarios where existing methods fall short.
\end{itemize}

\section{System Model} \label{sec:systemmodel}

\subsection{Network Model}

We consider a communication network described by a directed graph $\mathcal{G} \!\!=\!\! (\mathcal{V}, \mathcal{E})$, where $\mathcal{V}$ and $\mathcal{E}$ denote the set of nodes and links, respectively, $(i,j) \!\!\in\!\! \mathcal{E}$ denotes the link from node $i \!\in\! \mathcal{V}$ to node $j \!\in\! \mathcal{V}$, and $\rho_{i}^{-}$ and $\rho_{i}^{+}$ represent the incoming and outgoing set of neighbor nodes of node $i\in\mathcal V$, respectively. 

In order to support packet transmission, resource blocks (e.g., time-frequency blocks, optical wavelengths) can be allocated on each link $(i,j)\in\mathcal E$. 
We consider a time-slotted system with equal-sized slots $t \in \{0,1,\ldots\}$, and denote by $C^b_{ij}$ the capacity per resource block (in packets per slot) on link $(i,j)$, $X_{ij}^{max}$ the maximum number of available resource blocks on link $(i,j)$, and $C_{ij} = C^b_{ij}X_{ij}^{max}$ the link capacity of link $(i,j)$. 
We also denote by $e_{ij}$ the 
cost (e.g., power consumption) of operating a resource block on link $(i,j)$.

\subsection{Service Model}

The network can support multiple latency-sensitive service flows, or {\em commodities}, each requiring timely delivery of packets across specific source-destination pairs, represented by the set $\mathcal{C}$. 
QoS requirements are imposed by associating a {\em lifetime} or {\em time-to-live} (TTL) to packets of each commodity. 
A packet is considered {\em effective} if it has positive lifetime, and {\em outdated} otherwise. 
We define the {\em timely throughput} as the rate of effective packet delivery, i.e., the rate of packets delivered on time.

Packets of commodity $c \in \mathcal{C}$ are associated with (i) source node $s^{c} \in \mathcal{V}$, (ii) destination node $d^{c} \in \mathcal{V}$, 

(iii) initial lifetime $L^{c} \in \mathcal L = \{1,\dots,L_{max}\}$ with $L_{max}$ being the maximum possible lifetime, 
and (iv) {\em reliability} (timely throughput divided by the mean arrival rate) target $\delta^{c} \in [0,1]$.

In addition, 
the number of commodity-$c$ packets arriving at source node $s^{c}$ 
at time $t$ is denoted by a random variable $b^{c}(t)$, 
with a mean arrival rate of $\bar{b}^{c} = \mathbb{E} \big[ b^{c}(t) \big]$. We use $\bm{b}(t) \triangleq \big\{b^{c}(t), \forall c \in \mathcal{C}\big\}$ to denote vector of packet arrivals at time $t$.

\subsection{Network Control Decisions}

We consider network control policies  that perform (i) routing ({\em where} packets should go), (ii) scheduling ({\em when} packets should go), and (iii) resource allocation decisions. 

Routing and scheduling decisions are described by flow variables 
${\bf f}(t) \triangleq \big\{ f_{ij}^{(c,\ell)}(t), \forall (i,j) \in \mathcal{E}, \forall c \in \mathcal{C}, \forall \ell \in \mathcal{L} \big\}$, 
where $f_{ij}^{(c,\ell)}(t)$ denotes the number of packets of commodity $c$ and lifetime $\ell$ sent over link $(i,j)$ at time $t$.
For ease of notation, we also define $f_{i \rightarrow}^{(c,\ell)}(t) \triangleq \sum_{j \in \rho_{i}^{+}} f_{ij}^{(c,\ell)}(t)$ and $f_{\rightarrow i}^{(c,\ell)}(t) \triangleq \sum_{j \in \rho_{i}^{-}} f_{ji}^{(c,\ell)}(t)$ as the total outgoing flow from node $i$ to its outgoing neighbors of packets of commodity $c$, lifetime $\ell$, at time $t$, and the total incoming flow to node $i$ from its incoming neighbors, respectively.



To account for policies that can proactively drop packets from the network before lifetime expiry, we define packet-dropping variables
${\bf g}(t) \triangleq \big\{ g_{i}^{(c,\ell)}(t), \forall i \in \mathcal{V}, \forall c \in \mathcal{C}, \forall \ell \in \mathcal{L} \big\}$, where $g_{i}^{(c,\ell)}(t)$ denotes the number of commodity $c$, lifetime $\ell$ packets dropped by node $i$ at time $t$.

Finally, we define resource allocation variables ${\bf x}(t) \triangleq \big\{x_{ij}(t), \forall (i,j) \in \mathcal{E} \big\}$, where $x_{ij}(t)$ denotes the number of resource blocks allocated on link $(i,j)$ at time $t$, with $x_{ij}(t) \leq X_{ij}^{max}$.

\subsection{Lifetime-Based Queue Dynamics}

Upon arrival, packets build up on network queues according to their remaining lifetime and the commodity they belong to. 
We define queue variables ${\bf q}(t) \triangleq \{ q_{i}^{(c,\ell)}(t), \forall i \in \mathcal{V}, \forall c \in \mathcal{C}, \forall \ell \in \mathcal{L} \}$ where $q_{i}^{(c,\ell)}(t)$ denotes the queue backlog of node $i$ for packets of commodity $c$ and lifetime $\ell$ at time $t$.
Packets in the network experience progressive aging, which causes them to gradually deplete their lifetimes after every time slot.
Queue dynamics are given as
\begin{align}
    &q_{i}^{(c,\ell)}\!(t) \! = q_{i}^{(c,\ell \!+\! 1)}\!(t-1) \!\! - \!\! f_{i \rightarrow}^{(c,\ell \!+\! 1)}\!(t-1) \! - g_{i}^{(c,\ell+1)}\!(t-1) \nonumber \\ 
    & \hspace{1.2cm} + \! f_{\rightarrow i}^{(c,\ell \!+\! 1)}\!(t-1) \!  + \! b_i^{(c,\ell)}\!(t),
    \label{eq:lifetime-queue}
\end{align}
$\forall i \in \mathcal{V}, \forall c \in \mathcal{C}, \forall \ell \in \mathcal{L}, \forall t$, and we define $b_{i}^{(c,\ell)}(t)$ as
\begin{equation*}
  b_{i}^{(c,\ell)}(t) =
    \begin{cases}
      b^{c}(t) & \text{if $i = s^{c}, \ell = L^{c}$}\\
      0 & \text{otherwise}
    \end{cases}       
\end{equation*}

\begin{figure} 
    \centering
    \includegraphics[width=\columnwidth]{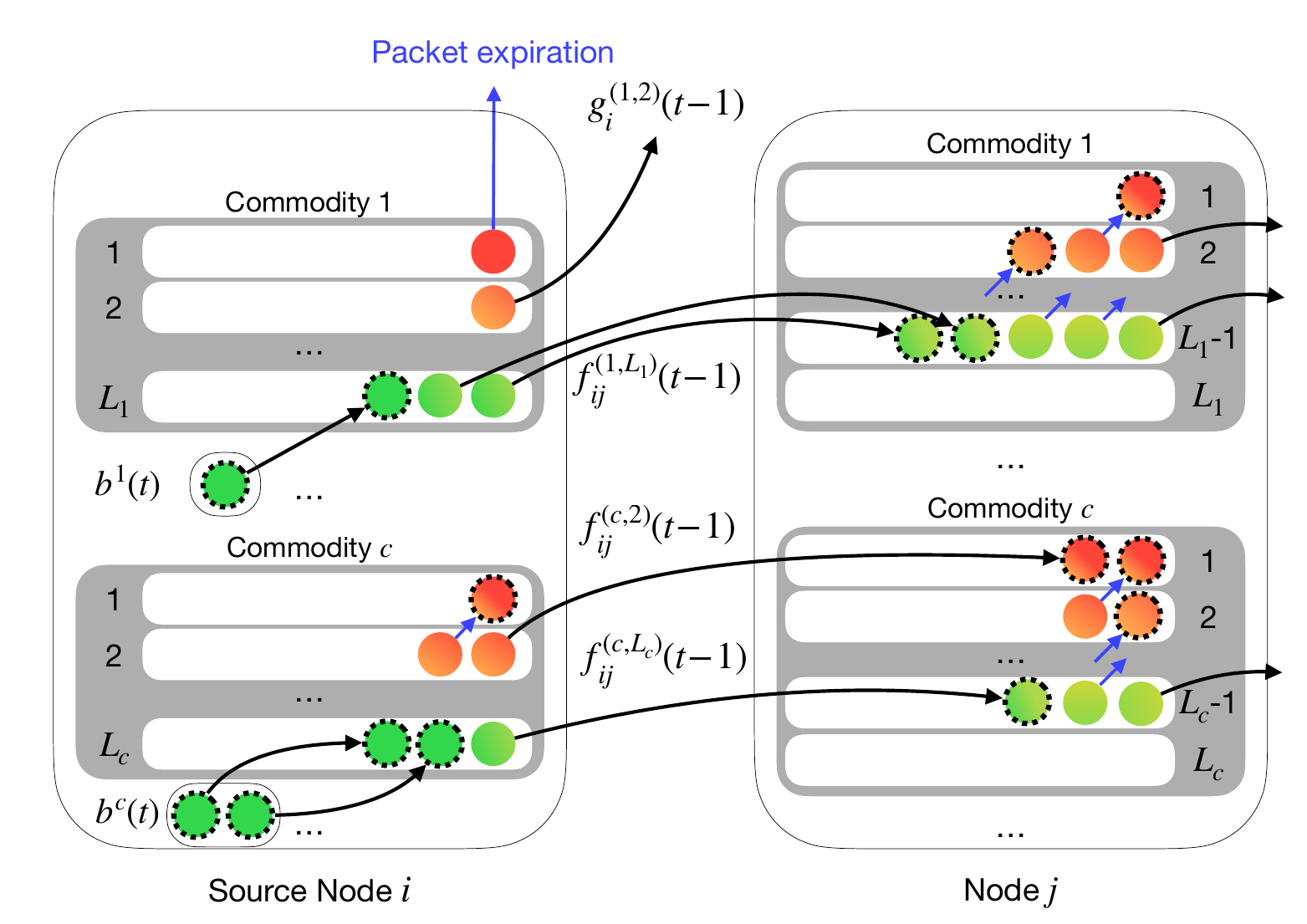}
    \caption{Illustration of lifetime-based queue dynamics: Packets turn from green to red as their lifetime decreases, blue arrows represent lifetime queue evolution, and packets without edges and with dashed edges indicate available packets at times $t-1$ and $t$, respectively.}
    \label{fig:lifetime_queue}
\end{figure}

We assume that {\em expired} (i.e., ineffective) packets are 
immediately dropped from the queue backlog, i.e.,
\begin{align}
    q_{i}^{(c,0)}(t) &= 0, &&\forall i \in \mathcal{V}, \forall c \in \mathcal{C}, \forall t,
    \label{eq:expired-packets}
\end{align}
and packets reaching their destination with positive remaining lifetime $\ell$ are immediately consumed, i.e.,
\begin{align}
    q_{d_{c}}^{(c,\ell)}(t) &= 0, && d_{c} \in \mathcal{V}, \forall c \in \mathcal{C}, \forall \ell \in \mathcal{L}, \forall t.
    \label{eq:packets-destination}
\end{align}
An illustration of the queueing dynamics is shown in Fig. \ref{fig:lifetime_queue}. 

\section{Problem Formulation} \label{sec:problem}

\textbf{Objective Function:}
The objective is to minimize the long-term average resource allocation cost,
\begin{align}
    \lim_{T \rightarrow \infty} \frac{1}{T} \sum_{t = 0}^{T-1} \sum_{(i,j) \in \mathcal{E}} \mathbb{E} [x_{ij}(t) e_{ij}]
\end{align}

\textbf{Reliability (Timely Throughput) Constraint:} Defining the timely throughput of commodity $c$ as the rate of {\em on-time packet delivery}, i.e., $\bar f_{\rightarrow d^{c}} \triangleq \mathlarger{\lim_{T \rightarrow \infty} \frac{1}{T} \sum_{t = 0}^{T-1} \sum_{\ell \in \mathcal{L}}} \mathbb{E} [f_{\rightarrow d^{c}}^{(c,\ell)}(t)], \forall c \in \mathcal{C}$, 
the reliability constraint imposes the timely throughput of commodity $c$ to surpass a prescribed {\em reliability} level $\delta^{c}$, i.e.,
\begin{align}
    \bar f_{\rightarrow d^{c}} -  \delta^{c} \bar{b}^{c} \geq 0, \hspace{0.25cm} &\forall c \in \mathcal{C}
    \label{eq:reliability_constraint}
\end{align}

\textbf{Availability Constraint:} The total outgoing flow (sent and dropped packets) cannot exceed the existing queue backlog:
\begin{align}
    f_{i\rightarrow}^{(c,\ell)}\!(t) \!+\! g_{i}^{(c,\ell)}\!(t) \!\leq\! q_{i}^{(c,\ell)}\!(t), ~ \forall i \in \mathcal{V}, \forall c \in \mathcal{C}, \forall \ell \in \mathcal{L}, \forall t
    \label{eq:availability_constraint}
\end{align}
\textbf{Capacity Constraint:} 
The total flow on link $(i,j)$ is limited by the allocated capacity, and the number of allocated resource blocks s is bounded by $X_{ij}^{max}$, i.e.,
\begin{align}
    &&& \sum_{\ell \in \mathcal{L}} \sum_{c \in \mathcal{C}} f_{ij}^{(c,\ell)}(t) \leq C^b_{ij}x_{ij}(t) & \forall (i,j) \in \mathcal{E}, \forall t \\
    &&& x_{ij}(t) \leq X_{ij}^{max} & \forall (i,j) \in \mathcal{E}, \forall t \label{eq:capacity_constraint}
\end{align}
\textbf{Problem Formulation:} The MDNC problem aims to find network control decisions that minimize resource allocation cost subject to queue dynamics, reliability, availability, and capacity constraints:
\begin{subequations}
\begin{align}
&\hspace{-0.23cm}\text{min} &&\hspace{-0.25cm} 
\lim_{T \rightarrow \infty} \frac{1}{T} \sum_{t = 0}^{T-1} \sum_{(i,j) \in \mathcal{E}} \mathbb{E} [x_{ij}(t) e_{ij}]\\
&\hspace{-0.23cm}\text{s.t.}  &&\hspace{-0.25cm} 
\bar f_{\rightarrow d^{c}} -  \delta^{c} \bar{b}^{c} \geq 0,
\hspace{2.4cm} \forall c \in \mathcal{C} \\ 
    &&& \hspace{-0.25cm}\text{Queues evolves by  \eqref{eq:lifetime-queue}-\eqref{eq:packets-destination}} \\
    &&&\hspace{-0.25cm} f_{i\rightarrow}^{(c,\ell)}\!(t) \!\!+\!\! g_{i}^{(c,\ell)}\!(t) \!\leq\! q_{i}^{(c,\ell)}\!(t),\forall i \!\in\! \mathcal{V},\forall c \!\in\! \mathcal{C},\forall \ell \!\in\! \mathcal{L},\forall t \\
    &&&\hspace{-0.25cm} \sum_{\ell \in \mathcal{L}} \sum_{c \in \mathcal{C}} f_{ij}^{(c,\ell)}(t) \leq C^b_{ij}x_{ij}(t) \hspace{0.8cm} \forall (i,j) \in \mathcal{E},  \forall t \\
    &&&\hspace{-0.25cm} x_{ij}(t) \leq X_{ij}^{max} \hspace{2.8cm} \forall (i,j) \in \mathcal{E}, \forall t \\
    &&&\hspace{-0.25cm} f_{ij}^{(c,\ell)}(t) \geq 0, ~g_{i}^{(c,\ell)}(t) \geq 0,~ x_{ij}(t) \in \mathbb{Z}^{+} 
\end{align}
\label{eq:control-problem} 
\end{subequations}
It is important to note that in network control problems that only require bounded average delays, such a requirement can be satisfied by imposing the stability of the underlying queuing system, which can in turn be satisfied by imposing flow conservation. This reduces the underlying network control problem to a stochastic optimization problem that admits efficient Lyapunov-based solutions\cite{neely2013dynamic}. However, in delay-constrained network control problems with lifetime-based queuing systems and associated packet drops, as in our MDNC problem, queue stability loses its purpose. As a result, the problem cannot be efficiently addressed via stochastic optimization methods, even though heuristic solutions have been designed as in~\cite{cai2022ultra}. However, the MDNC problem can be cast 
as a CMDP, which incorporates delay constraints directly into network decisions. This approach is better suited to handle the dynamic and time-sensitive nature of packet lifetimes in NextG networks, as illustrated in the following section.

\section{CDRL for Dynamic Network Control}

We now show how the MDNC problem stated in Section \ref{sec:problem} can be modeled as a CMDP and solved via CDRL. 

A CMDP is characterized by its state transition probability distribution $\mathbb{P}({\bf s}(t), r(t) \given {\bf s}(t-1), {\bf a}(t-1))$, representing the probability of an agent reaching state ${\bf s}(t)$ and obtaining reward $r(t): \mathcal{S} \times \mathcal{A} \rightarrow \mathbb{R}$ after taking action ${\bf a}(t-1) \in \mathcal{A}$ in state ${\bf s}(t-1) \in \mathcal{S}$, 
where reward $r(t)$ quantifies the effectiveness of action ${\bf a}(t-1)$ in state ${\bf s}(t-1)$. The agent selects actions based on a policy $\pi$, where $\pi({\bf s}(t))={\bf a}(t) \in \mathcal{A}_{{\bf s}(t)}$. The goal of the agent is to maximize the {\em return} (i.e., expected sum of discounted rewards), defined as $\mathbb{E}_{s} [ \sum_{t = 0}^{\infty} \gamma^{t} r(\pi({\bf s}(t)), {\bf s}(t)) ]$, where $\gamma \in [0,1]$ is the discount factor.

\textbf{State Space:} 
The state of node $i$ at time $t$ includes corresponding exogenous packet arrivals and queue backlogs, given by ${\bf s}_{i}(t) = \{b_{i}^{(c,\ell)}(t), q_{i}^{(c,\ell)}(t): i \in \mathcal{V}, \forall c \in \mathcal{C}, \forall \ell \in \mathcal{L} \}$, and the overall network state 
becomes ${\bf s}(t) = \{ {\bf s}_{i}(t): \forall i \in \mathcal{V}\}$. 

\textbf{Action Space:} Network actions consist of resource allocation variables ${\bf x}(t)$, and packet routing and scheduling decisions encoded into variables ${\bf f}(t)$ and ${\bf g}(t)$. The network action at time $t$ is denoted by ${\bf a}(t) = \big\{ {\bf x}(t), {\bf f}(t), {\bf g}(t) \big\}$, 
where the action space $\mathcal{A}_{{\bf s}(t)}$ depends on the network state ${\bf s}(t)$, i.e.,
\begin{align*}
    &\mathcal{A}_{{\bf s}(t)} \!\!=\!\! \Big\{ \! x_{ij}\!(t), f_{ij}^{(c,\ell)}\!(t), g_{i}^{(c,\ell)}\!(t)\!:\! f_{i\rightarrow}^{(c,\ell)}(t) \!+\! g_{i}^{(c,\ell)}\!(t) \!\leq\! q_{i}^{(c,\ell)}(t), \\ 
    & 0 \!\leq\!\! \sum_{c \in \mathcal{C}}\! \sum_{\ell \in \mathcal{L}} \!f_{ij}^{(c,\ell)}\!(t) \!\leq\! C_{ij}^{b}x_{ij}\!(t), 0 \!\leq\! x_{ij}\!(t) \! \leq \! X_{ij}^{max}, x_{ij} \! \in \! \mathbb{Z}^{+} \! \Big\},
\end{align*}

$\forall (i,j)\!\! \in \!\! \mathcal{E},\forall \ell \! \in \! \mathcal{L},\forall i \! \in \! \mathcal{V},\forall c \! \in \! \mathcal{C}$. 

\textbf{CDMP-Based Formulation:} 
In our formulation, we consider weakly communicating MDPs, which is a common assumption in the study of infinite-horizon RL. While traditional MDP objectives typically focus on maximizing the expected sum of discounted rewards, the MDNC problem aims to minimize the infinite-horizon average expected cost. Theorem 1 in \cite{wei2020model} establishes that even under the weaker assumption of weakly communicating MDPs, an optimal policy can still be learned by optimizing the infinite-horizon average expected reward instead of relying on discounted objectives. Therefore, in this paper, we shift our focus on the discounted average reward framework as the basis for our CMDP formulation. Based on this, we define the infinite-horizon discounted average as
\begin{align}
{\widehat m}_0(\pi) \triangleq  \sum_{t = 0}^{\infty} \gamma^{t} \mathbb{E}_{s} \left[m_0({\bf s}(t), {\bf a}(t)) \right],
\label{eq:cmdp-m0}
\end{align}
where $m_0({\bf s}(t), {\bf a}(t)) \triangleq \sum_{(i,j) \in \mathcal{E}} x_{ij}(t) e_{ij}$ \footnote{Recall that $x_{ij}(t) \in \mathcal{A}_{{\bf s}(t)}$, and hence depends on the chosen policy $\pi({\bf s}(t))$ and the state ${\bf s}(t)$.}.

Using \eqref{eq:cmdp-m0} as the objective function and rewriting the constraints in \eqref{eq:control-problem} in terms of state and actions, the CDMP-based MDNC problem becomes: 
\begin{subequations}
\begin{align}
&\text{min} && {\widehat m_0}(\pi) \\
&\text{s.t.} && {\widehat m^{c}}(\pi) \geq 0, & \forall c \in \mathcal{C} \\
&&& \pi({\bf s}(t)) \in \mathcal A_{{\bf s}(t)}, & \forall t 
\end{align}
\label{eq:primalhat} 
\end{subequations}
where $\widehat m^{c}(\pi) \!\triangleq\! \mathlarger{\sum_{t = 0}^{\infty} \sum_{\ell \in \mathcal{L}}} \gamma^t \mathbb{E} [f_{\rightarrow d^{c}}^{(c,\ell)}(t)] \!-\! \delta^{c} \bar{b}^{c}, \hspace{0.2cm} \forall c \in \mathcal{C}$.

\subsection{Dual Subgradient Algorithm}
As shown in~\cite{crl}, the dual subgradient approach can be utilized to find an optimal policy for problem \eqref{eq:primalhat} iteratively. 
The Lagrangian for the problem can be written as 
\begin{align}
    \mathcal{L}(\pi,\bm{\lambda}) &\triangleq {\widehat m_0}(\pi) - \sum_{c \in \mathcal{C}} \lambda^{c} {\widehat m^{c}}(\pi),
\end{align}
where $\bm{\lambda} = \{ \lambda^{c}, \forall c \in \mathcal{C} \}$ is the vector of Lagrangian multipliers. Then, the dual problem becomes
\begin{align}
&\max_{\bm{\lambda}} \inf_\pi && \mathcal{L}(\pi,\bm{\lambda})\\
&&& \lambda^{c} \geq 0  & \forall c \in \mathcal{C} \notag
\end{align}

Since the dual problem is convex in $\bm{\lambda}$, we can use the dual subgradient method to find optimal $\lambda^{c}$'s iteratively. 
At each iteration $k$, the primal (i.e., policy)
updates 
are performed as
\begin{align}
& \pi_{k+1} = \argmin_{\pi} \, \mathcal{L}(\pi, \bm{\lambda}_{k}) .
\end{align}
Subsequently, using the updated policy, the dual (i.e., Lagrange multipliers) updates at iteration  $k$ are performed as
\begin{align}
\lambda^{c}_{k+1} &= \Big[ \lambda^{c}_{k} - \eta^{c} \, \widehat m^{c} \big( \pi_{k+1} \big) \Big]^+, & \forall c \in \mathcal{C}
\end{align}
where $\eta^{c}$ is the dual learning rate for commodity $c$ and $[.]^{+}$ is the projection operator onto the positive orthant.
\subsection{DRL-based Primal-Dual Update}
The work in \cite{crl} showed that there is zero duality gap between primal and dual problems, enabling the dual subgradient algorithm to find the optimal policy. They also showed that while in setups with high-dimensional state-action spaces, memory and computational requirements to find the optimal policy grow exponentially, the policy 
can be learned through experience (e.g., via RL). 

Indeed, CDRL provides methods to learn optimal policies for CMDP-defined problems, where certain constraints must be met while maximizing the return. To this end, the policy is parameterized by a vector $\bm{\theta} \in \mathbb{R}^{w}$ with number of parameters $w$, learned by a DRL algorithm. The instantaneous reward for the parameterized policy becomes
\begin{align}
r_{\bm{\lambda}}({\bf s}(t),{\bf a}(t)) &\triangleq - m_0({\bf s}(t),{\bf a}(t)) \nonumber \\
& \hspace{0.5cm} + \sum_{c \in \mathcal{C}} \lambda^{c} m^{c}({\bf s}(t),{\bf a}(t)) 
\label{eq:instantaneous_reward}
\end{align}
Then, the return can be written as 
\begin{align}
    & \mathbb{E}_{s} \bigg[ \sum_{t = 0}^{\infty} \gamma^{t} r_{\bm{\lambda}}({\bf s}(t),{\bf a}(t)) \bigg] \! = \nonumber \\
    & \sum_{t = 0}^{\infty} \!\! \gamma^{t}\mathbb{E}_{s} \!\! \bigg[ \!\! - \! m_0( {\bm s}(t),\!{\bf a}(t)) \! + \!\! \sum_{c \in \mathcal{C}} \! \lambda^{c} m^{c}\!({\bm s}(t),\!{\bf a}(t)) \! \bigg] \nonumber \\
    & \hspace{4.45cm} = - \mathcal{L}(\pi_{\bm{\theta}},\bm{\lambda}) 
\end{align}

This shows that the return can be expressed in terms of the parameterized Lagrangian, enabling the solution to the dual problem (i.e., minimizing the Lagrangian) to be obtained through a parameterized policy $\pi_{\theta}$ via CDRL methods \cite{crl}.

Using the parameterized Lagrangian, the DRL algorithm performs the policy parameters update at iteration $k$ as
\begin{align}
    \bm{\theta}_{k+1} &\approx \argmin_{\bm{\theta}} \mathcal{L}(\bm{\theta}, \bm{\lambda}_{k}),
\end{align}
Dual multipliers can be updated using the updated policy as
\begin{align}
& \lambda^{c}_{k+1} = \left[ \lambda^{c}_{k} - \eta^{c} \, \widehat m^{c}(\pi_{\bm{\theta}_{k+1}}) \right]^+, & \forall c \in \mathcal{C}
\label{eq:dual_lambda_update}
\end{align}
At each dual subgradient algorithm iteration $k$, we perform a primal update via DRL over $V$ training episodes and a dual update. Each training episode lasts $T$ steps, 
hence primal 
and dual 
updates are performed 
at every $TV$ steps. The pseudocode of CDRL-NC is given in Algorithm \ref{algo}.

\begin{algorithm}
\bf{Input:} $\delta^{1},\ldots,\delta^{|\mathcal{C}|}, \lambda^{1}_{0},\ldots,\lambda^{|\mathcal{C}|}_{0}$ \\
Initialize: $\bm{\theta}_{0}$\;
    \For{\text{iteration} k = 0,1,\ldots}{
        \For{\text{episode} v = 0,1,\ldots, V}{
            \For{\text{step} t = 0,1,\ldots, T}{
                \hspace{-0.1cm}{\normalfont Observe ${\bf{s}}(t)$, take action ${\bf{a}}(t)$, get reward $r_{\bm{\lambda}}({\bf s}(t),{\bf a}(t))$}\;
            }
        }
        {\normalfont Sample from observations}\;
        {\normalfont Perform primal update: $\bm{\theta}_{k+1}$ via DRL}\footnotemark \;
        \For{$c = 1,\ldots, |\mathcal{C}|$}{
            {\normalfont Dual update: $\lambda^{c}_{k+1} = \left[ \lambda^{c}_{k} + \eta^{c} \, \widehat m^{c}(\pi_{\bm{\theta}_{k+1}}) \right]^+$}\;
        }
        \For {\normalfont{\textbf{each}} $k \% K == 0 $}{
            {\normalfont Save model if model checkpoint is satisfied}
        }
    }
 \caption{CDRL-NC}
 \label{algo}
\end{algorithm}
\footnotetext{Note that this step can be performed for multiple agents simultaneously.}
\subsection{CDRL-NC Agents Design}

We adopt the actor-critic based multi agent deep deterministic policy gradient (MADDPG) algorithm as the DRL method, allowing to train multiple agents in an offline manner serving different purposes \cite{lowe2017multi}. During the training phase, each agent as an {\em actor} learns a policy, while the global {\em critic} aims to maximize the return.

We focus on policies with centralized routing and distributed scheduling (akin to software-defined networking \cite{stampa2017deep}), where a single routing agent makes routing decisions for all commodities,  whereas scheduling agents at every node take local scheduling actions. The scheduling agent at node $i \in \mathcal{V}$ learns a policy which uses the local information $s_{i}(t)$, and the routing agent learns a policy using the network state ${\bf s}(t)$. Upon arrival at source node $s^c$, packets of commodity $c$ are assigned to a path $p\in\mathcal P^c \subset \mathcal{P}$, with $\mathcal{P}^{c}$ denoting the set of  feasible paths for commodity $c$, and $\mathcal{P} = \bigcup_{c \in \mathcal{C}} \mathcal{P}^{c}$. Given that each packet gets assigned to a path at the source, we denote the number of packets assigned to path $p$, with lifetime $\ell$, held at node $i$ at time $t$ as $q_{i}^{(p,\ell)}(t)$, with ${\bf q}(t) = \{ q_{i}^{(p,\ell)}(t), \forall i \in \mathcal{V}, \ell \in \mathcal{L}, \forall p \in \mathcal{P} \}$, now denoting the {\em path-based} network queue backlog.

The \underline{\textbf{routing agent}} assigns a path to each packet upon its arrival to the network. 

At each time slot, 
the routing agent's {\em actor} observes the path-based queue backlog ${\bf q}(t)$ and packet arrivals ${\bf b}(t)$, and for each commodity $c$, outputs a $|\mathcal{P}^{c}|$ dimensional probability vector. Each entry in this vector, denoted by $\mathbb{P}^{c}\{ p \}$, specifies the proportion of packets of commodity $c$ to be routed along path $p \in \mathcal{P}^{c}$. Accordingly, the number of packets of commodity $c$ assigned to path $p$ at time $t$ is given by $\lfloor b^{c}(t) \mathbb{P}^{c}\{ p \} \rfloor$.

The \underline{\textbf{scheduling agent}} {\em actor} at node $i$ observes the {\em aggregate} local state for path $p$, given by $q_{i}^{p}(t) = \sum_{\ell = 1}^{|\mathcal{L}|}q_{i}^{(p,\ell)}(t)$, and outputs a three dimensional probability vector $\mathbb{P}^{p}_{i}(x)$, where $x$ is mapped to packet ``send", ``drop", or ``hold" actions. Based on this, the number of packets sent (`s') and dropped (`d') for path $p$ by node $i$ at time $t$ are computed as
\begin{align}
S_{i}^{p}(t) = \lfloor\mathbb{P}_{i}^{p}(\text{`\text{s}'})q_{i}^{p}(t)\rfloor ~,~ D_{i}^{p}(t) = \lfloor\mathbb{P}_{i}^{p}(\text{`\text{d}'})q_{i}^{p}(t)\rfloor,
\end{align}
which are mapped to network-level forwarding and dropping decisions, ${\bf f}(t)$ and ${\bf g}(t)$. Starting with packets in $q_{i}^{p}(t)$ with the lowest lifetime, the agent at node $i$ drops $D_{i}^{p}(t)$ packets, sends $S_{i}^{p}(t)$ packets along path $p$, and retains the rest. The total number of packets sent over link $(i,j)$ at time $t$ is given by $S_{ij}(t) = \sum_{p \in \mathcal{P}_{ij}} S_{i}^{p}(t)$, where $\mathcal{P}_{ij}$ is the set of paths using link $(i,j)$. Then, the number of resource blocks allocated to link $(i,j)$ at time $t$ can be computed as 
\begin{align}
    x_{ij}(t) = \biggl\lceil \frac{S_{ij}(t)}{C_{ij}^{b}} \biggr\rceil.
\end{align}
 Note that  this decision-making process deliberately simplifies the observation space of scheduling agents by relying on aggregate path-based queue backlogs, $q_{i}^{p}(t)$. 
    Lifetime information is then used by a heuristic rule that favors sending packets more likely to reach their destinations. This hybrid design reduces inference complexity and promotes scalable, low-overhead scheduling decisions while striking a good balance between efficiency and performance. 
    Investigating the complexity-performance tradeoff of more fine-grained scheduling policies that incorporate lifetime information in the observation space is part of our ongoing studies.

Finally, the {\em critic} observes the network state ${\bf s}(t)$ and actions ${\bf a}(t)$ at time $t$ to estimate the expected sum of discounted reward for the state-action pair. Note that the reward $r(t)$ is used by all agents. 
\subsection{Complexity Analysis}
We now analyze the inference complexity to perform routing and scheduling decisions, as well as the communication overhead of the proposed CDRL-NC algorithm. Let the neural networks used by each agent be modeled as multi-layer perceptrons (MLPs) with two hidden layers containing $n_1$ and $n_2$ neurons, respectively. For the centralized routing agent, the input consists of the network queue backlog ${\bf q}(t)$ and packet arrivals ${\bf b}(t)$, while the output is a path assignment for each arriving packet, and its inference complexity is given by $\mathcal{O}((|\mathcal{V}||\mathcal{P}||\mathcal{L}|)+|\mathcal{C}|) n_{1} + n_{1} n_{2} + n_{2} |\mathcal{P}|)$. For the distributed scheduling agents, each node $i$ observes only its local aggregate queue states $q_i^p(t)$ as input and outputs flow decision variables for each path $p$. Therefore, the inference complexity of each scheduling agent can be written by $\mathcal{O}(|\mathcal{P}| n_{1} + n_{1} n_{2} + n_{2} 3|\mathcal{P}|)$.

In contrast to Lyapunov-based approaches, there is no communication overhead during packet scheduling since only the local state is used for scheduling decisions. On the other hand, the centralized router requires local states from each node to make routing decisions, incurring the same communication overhead as that of existing centralized routing 
policies, such as UMW~\cite{sinha2017optimal}. 

\subsection{Practical Considerations of CDRL-NC}
Convergence of both $\bm{\lambda}$ and $\bm{\theta}$ can be time-intensive in some setups. To address this, we propose pseudo-convergence model checkpoint criteria to capture and save the best-performing model. 

The model at iteration $k$ is said to show stable and satisfactory performance, and can be saved if
\begin{subequations}
\begin{align}
    {\widehat m^{c}}(\pi_{\bm{\theta}_{k}}) &\geq 0 ~~ \forall c,  \\
    \bm{\sigma}^{K}_{\bm{\lambda}} &< \sigma_{\text{Thr}}, \\
    \bar{\bf r}^{K} &> \bar{\bf r}^{\text{max}},
\end{align}
\end{subequations}
where $\bm{\sigma}^{K}_{\bm{\lambda}}$ denotes the standard deviation of $\bm{\lambda}$ over last $K$ policy iterations, with $\sigma_{\text{Thr}}$ as a predefined threshold, $\bar{\bf r}^{K}$ is the average reward over length-$K$ sliding window, with $\bar{\bf r}^{\text{max}}$ being the maximum average reward observed in prior windows.

When the timely throughput conditions are satisfied, $\bm{\lambda}$ shows low variability, and a high reward is achieved, the model is saved to indicate near-convergence performance.

\section{Experimental Results}
We now present experiments to evaluate the performance of CDRL-NC across different settings. We adopt the MADDPG algorithm with discount factor $\gamma = 0.97$, and we apply $\epsilon$-greedy strategy with a policy exploration probability of $\max((0.99)^{k},0.01)$ at iteration $k$ in training to balance exploration and exploitation\cite{sutton2018reinforcement}. To further explore the state-action space, we divide policy training into {\em train} and {\em improve} pheases, prior to policy evaluation, during which the $\epsilon$-greedy strategy is applied at each phase to enable additional policy exploration. After training and improve phases, we evaluate the performance of the best performing policy in the test phase. 
Agents' actors and the critic have MLP of two hidden layers with $64$ neurons at each hidden layer, and Adam optimizer is applied with a  learning rate of $10^{-3}$ \cite{kingma2014adam}. During simulations, the cost $m_{0}$ and timely throughput values, $m^{c}$,  are normalized within $[0,1]$ to maintain comparable orders of instantaneous cost and timely throughput values \cite{engstrom2020implementation}.

\begin{figure} 
    \centering
    \includegraphics[width=\columnwidth]{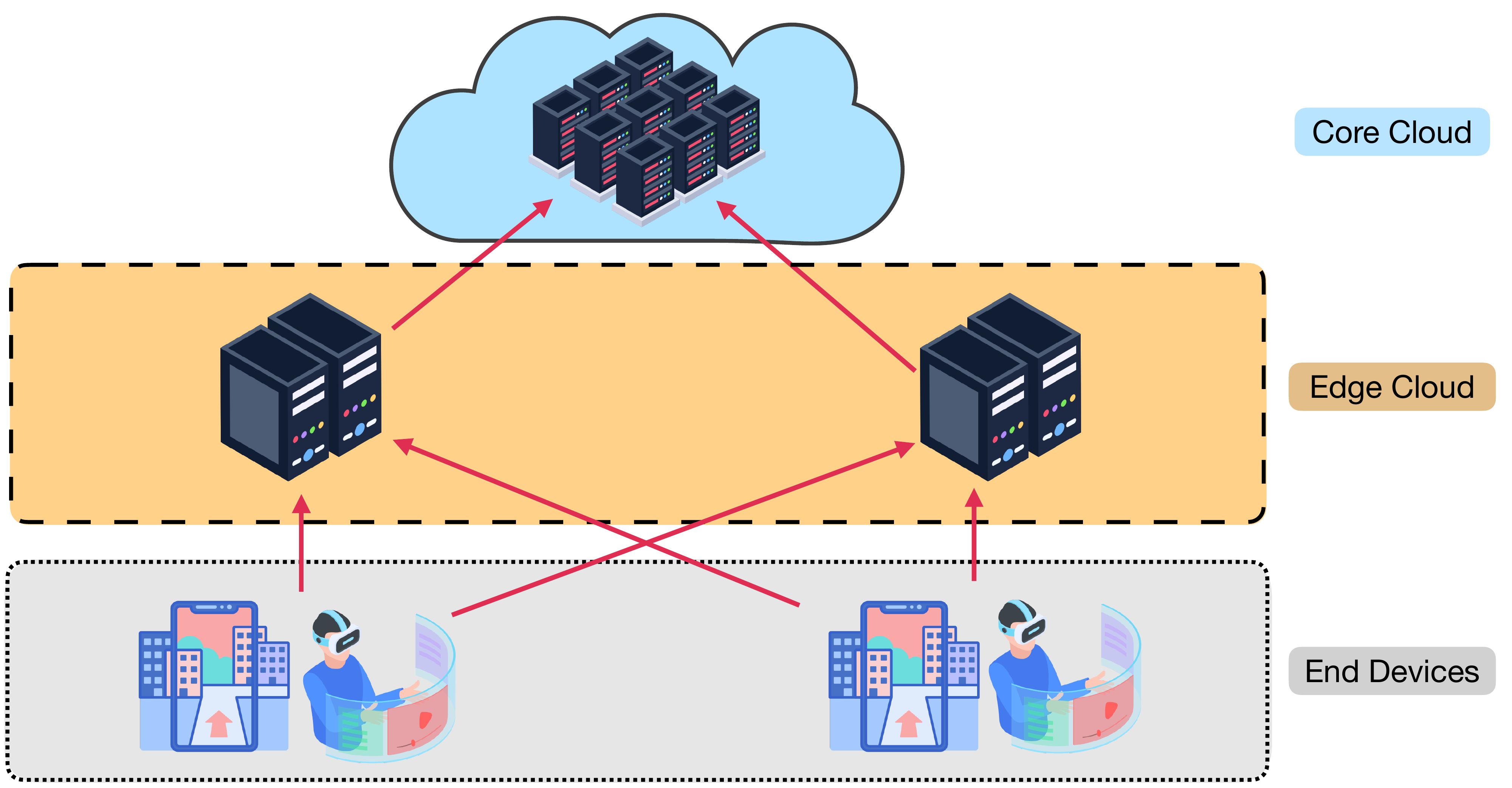}
    \caption{Illustration of an edge network topology.} \label{fig:networks}
\end{figure}

We perform experiments to evaluate the performance of CDRL-NC with competing algorithms, including BP, which uses distributed routing and scheduling, and UMW, which employs centralized routing and distributed scheduling \cite{tassiulas1990stability, sinha2017optimal}. We employ the edge network architecture with two edge cloud servers in between, similar to what is presented in \cite{alsabah20216g}, shown in Fig. \ref{fig:networks}.
The packet arrivals follow a Poisson distribution, links have a capacity of 10 packets per slot with equal link costs, i.e.,  $e_{ij}=1, ~ \forall (i,j) \in \mathcal{E}, \forall t$.
We set the dual learning rate to $\eta^{c} = 0.005$ and initial Lagrangian multipliers to $\lambda^{c}_{0} = 1.25 \sqrt{\bar{b}^{c} \delta^{c}}, c \in \{1,2\}$ to make the algorithm adaptive to varying timely throughput constraints. This approach allows commodities with more stringent constraints to prioritize timely throughput over minimizing total cost early in the experiment. After the commodity $c$ meets $m^{c}$, its $\lambda^{c}$ decreases via dual updates, reducing the weight corresponding to the throughput, and shifting the focus of instantaneous reward to the total cost rather than achieving the reliability target.


\begin{figure} [htbp]
    \centering
    \includegraphics[width=\columnwidth]{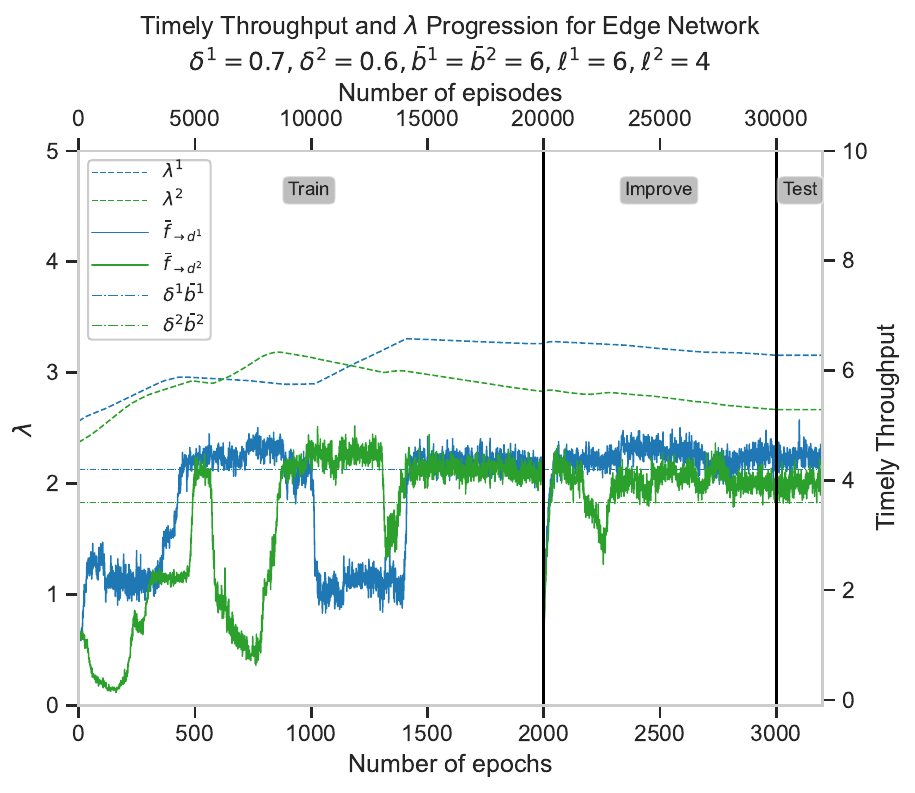}
    \caption{Progression of $\bm{\lambda}$ and average timely throughput of commodities as CDRL-NC agents train for the edge network. Solid curves represent the instantaneous timely throughput whose throughput constraints are marked with horizontal dotted dashed lines, and dashed curves represent the instantaneous $\bm{\lambda}$ values. When the timely throughput is low, $\bm{\lambda}$ values increase in order to prioritize timely throughput more in the reward function on Eq. \eqref{eq:instantaneous_reward}. As the timely throughput targets are satisfied, the $\bm{\lambda}$ values become more stable since the rightmost term on Eq. \eqref{eq:dual_lambda_update} gets closer to zero. } \label{fig:lambda_rel_diamond}
\end{figure}

We simulate CDRL-NC on the edge network where packets are routed and sent from two end devices to the core cloud node with reliability constraints $\delta^{1} = 0.7$, $\delta^{2} = 0.6$, and initial lifetimes $L^{1} = 6$, $L^{2} = 4$, respectively, and with episode length $T = 20$ time steps. The experiments include 20000 training episodes, 10000 improve episodes, and 2000 test episodes. In Fig. \ref{fig:lambda_rel_diamond}, when $\bar{b}^{1} = \bar{b}^{2} = 6$, we illustrate the progression of throughput and $\lambda^{c}$'s throughout training. Dashed lines represent $\lambda^{c}$ values for both commodities, while solid lines show their throughput, and horizontal lines indicate their corresponding timely throughput targets. As agents proceed into the improve phase, the throughput remains above the threshold, and $\lambda^{c}$ values gradually decrease, approaching an equilibrium point. In the test phase, the performance of the best model via model checkpoint is evaluated without further policy iterations.

\begin{figure} [htbp]
    \centering
    \captionsetup{width=\columnwidth}
    \begin{subfigure}[b]{1\columnwidth}  
        \centering
        \includegraphics[width=\columnwidth]{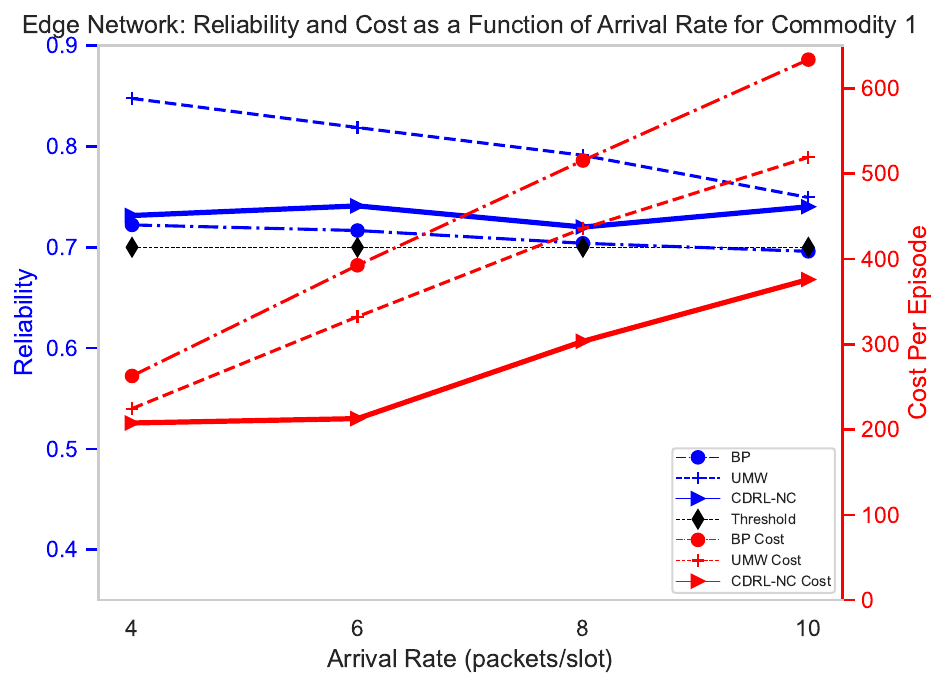}
        \caption{Commodity 1 of the edge network.}
    \end{subfigure}
    
    \begin{subfigure}[b]{\columnwidth}
        \centering
        \includegraphics[width=\columnwidth]{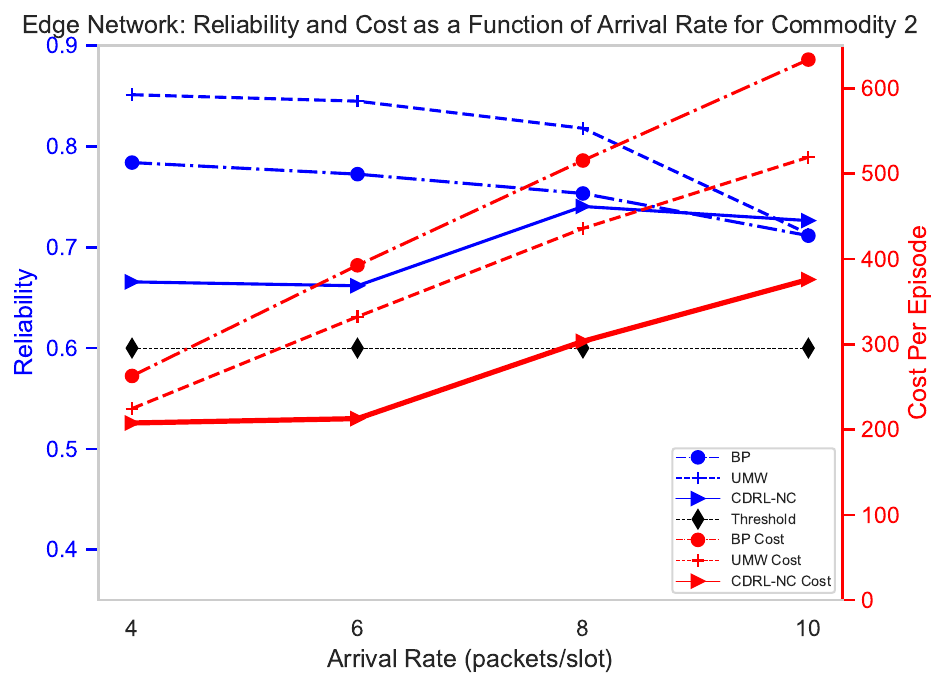}
        \caption{Commodity 2 of the edge network.}
    \end{subfigure}
    \caption{Reliability and cost per episode of two commodities in the edge network for reliability targets $\delta^{1} = 0.7$, $\delta^{2} = 0.6$ (marked with horizontal dashed black line) and initial lifetimes $L^{1} = 4$. The arrival rates are kept the same for both commodities. Even though all approaches are able to stay above the reliability targets, CDRL-NC is able to perform timely packet delivery with less cost than both BP and UMW. When $\bar{b}^{1} = \bar{b}^{2} = 10$, CDRL-NC is able to satisfy the reliability targets while BP fails to meet the reliability target for commodity 1.}
    \label{fig:rel_cost_diamond}
\end{figure}

Fig. \ref{fig:rel_cost_diamond}, presents the reliability levels, $\frac{f_{\rightarrow d^{c}}}{\bar{b}^{c}}$, reliability thresholds $\delta^{c}\bar{b}^{c}$, and network resource allocation cost per episode values for various commodities and arrival rates after policy training. In low arrival rate regime, all algorithms satisfy the timely throughput constraint, however, CDRL-NC consistently achieves a lower cost per episode. As the arrival rate increases, BP starts to fail achieving the reliability constraint for commodity 1 when $\bar{b}^{1}=\bar{b}^{2}=10$. UMW consumes lower cost than BP but CDRL-NC still satisfies the reliability constraints at a much lower cost.

\section{Conclusion}
In this paper, we have formulated the MDNC problem, where services are delay-sensitive and need to satisfy a prescribed reliability level, as a CMDP. To solve this problem, we have proposed a dual subgradient-based CDRL framework that learns cost and latency efficient routing and scheduling policies. Through network simulations, we have demonstrated that the proposed CDRL-NC algorithm converges to feasible policies under various settings, effectively meeting target reliability levels while achieving significantly lower cost than alternative approaches. Ongoing investigations include robustness analysis of CDRL-NC under different network topologies and service settings, as well as investigating various agent designs with different observation and action spaces, and compare their inference complexity-performance tradeoffs. 

\bibliography{cdrl_conference_references}
\bibliographystyle{ieeetr}

\end{document}